\documentclass{sf2a-conf2012}
\usepackage{graphicx}
\usepackage{hyperref}
\usepackage[cyr]{aeguill}
\usepackage{epstopdf}

\begin{document}

\TitreGlobal{SF2A 2012}


\title{The electromagnetic calorimeter of the AMS-02 experiment}

\runningtitle{The AMS-02 ECAL}

\author{M. Vecchi}\address{Academia Sinica, 128 Academia Road, Section 2, Nankang, Taipei 115, Taiwan (R.O.C.), and National Central University, No.300, Jhongda Rd., Jhongli City, Taoyuan County 32001, Taiwan (R.O.C.)}


\author{L. Basara}\address{LAPP, Universit\'e de Savoie, CNRS/IN2P3, Annecy-le-Vieux, France}
\author{G. Bigongiari}\address{Universita' degli Studi di Siena, 53100 Siena, Italy}
\author{F. Cervelli}\address{Istituto Nazionale di Fisica Nucleare, Sezione di Pisa, 56100 Pisa, Italy}
\author{G. Chen}\address{Institute of High Energy Physics, CAS, Beijing, 100049, China}
\author{G. M. Chen, $^5$}
\author{H. S. Chen, $^5$}
\author{G. Coignet, $^2$}
\author{S. Di Falco, $^4$}
\author{S. Elles, $^2$}
\author{A. Fiasson, $^2$}
\author{D. Fougeron, $^2$}
\author{G. Gallucci, $^4$}
\author{C. Goy, $^2$}
\author{M. Incagli, $^4$}
\author{R. Kossakowski, $^2$}
\author{V. Lepareur, $^2$}
\author{Z.H. Li, $^5$}
\author{M. Maire, $^2$}
\author{M. Paniccia, $^2$}
\author{F. Pilo, $^4$}
\author{S. Rosier-Lees, $^2$}
\author{X.W. Tang, $^5$}
\author{C. Vannini, $^4$}
\author{J.P. Vialle, $^2$}
\author{H.L. Zhuang $^5$}
\setcounter{page}{237}
\maketitle
\begin{abstract}
The electromagnetic calorimeter (ECAL) of the AMS-02 experiment is a 3-dimensional sampling calorimeter, made of lead and scintillating fibers.
The detector allows for a high granularity, with 18 samplings in the longitudinal direction, and 72 sampling in the lateral direction. 
The ECAL primary goal is to measure the energy of cosmic rays up to few TeV, however, thanks 
to the fine grained structure, it can also provide the separation of positrons from protons, in the GeV to TeV region. 
A direct measurement of high energy photons with accurate energy and direction determination can also be provided.
\end{abstract}
\section{Introduction}
\begin{figure}[ht!]
 \centering
 \includegraphics[width=0.25\textwidth,clip]{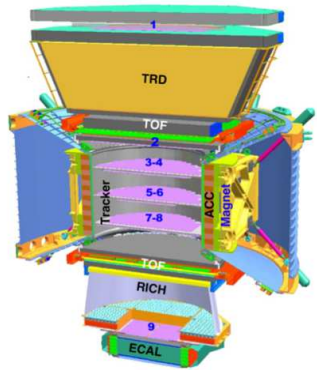} 
  \caption{Sketch of the AMS-02 detector.}
  \label{mvecchi:fig1}
\end{figure}

The Alpha Magnetic Spectrometer (AMS-02) (see~\cite{Aguilar}) is a large acceptance particle physics detector, operating onboard the International Space Station (ISS) since May 19th 2011. Since the launch, the detector has collected more than 20 billion events.
The main goals of AMS-02 are the search for antimatter and dark matter, and the precise measurements of cosmic rays composition and flux. Moreover, AMS-02 
could provide information on gamma rays in the GeV to TeV range.\\
The apparatus is composed of several subdetectors, arranged as in figure \ref{mvecchi:fig1}.
\begin{itemize}
\item a Transition Radiation Detector (TRD), equipped with 20 layers, each one using polypropylene/polyethylene as radiator and Xe/Co2 filled proportional wire straw tubes. This detector is used to complement the ECAL in particles identification, up to few hundred GeVs.
\item a Time of Flight (TOF), made of 4 layers of plastic scintillators, providing timing information and the trigger for the experiment.
\item a Magnetic Spectrometer, made of double sided silicon strips, arranged on 9 layers, to measure the sign of the electric charge and the rigidity (momentum/charge) of cosmic rays. The measured quantity is the deflection, i.e. the inverse of the rigidity, for particles in a 0.15 T magnetic field, provided by a permanent magnet.
\item an Anti-Coincidence scintillator Counter (ACC) system surrounds the tracker planes installed within the magnet volume. The ACC system provides 
the veto signal to the trigger, in order to reject multi-particle events, generated in the interaction of cosmic rays entering the detector through the sides.
\item a Ring Imaging Cherenkov Detector (RICH), to estimate cosmic rays velocity and charge with high accuracy.
\item an electromagnetic calorimeter (ECAL), to measure particles energy, for the electron/hadron separation and for the photon detection.
\end{itemize} 
\section{The ECAL detector }
\begin{figure}[ht!]
 \centering
 \includegraphics[width=0.3\textwidth,clip]{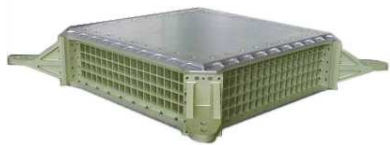}      
  \caption{The Electromagnetic Calorimeter honeycomb structure, before the final integration. PMTs are installed on the squared holes.}
  \label{vecchi:fig2}
\end{figure}

The ECAL (see~\cite{Cadoux}) is a lead-scintillating fibers sampling calorimeter, in which particles crossing the active volume produce light collected by photomultipliers (PMTs). 
The structure has been designed to have a high granularity (one radiation lenght in the longitudinal direction, half a Moli\`ere radius in the lateral direction) and to
maximize the ratio between radiation and interaction lenght. The resulting composite structure has a 
lead-fiber-glue volume composition of 1:0.57:0.15 cm$^3$, as well as an average density of 6.8 g/cm$^3$.
The active part of the detector has 
a total weight of approximately 638 kg and a total active size of 68.5~$\times$~68.5~$\times$~16.7~cm$^3$~(figure~\ref{vecchi:fig2}).
The total depth of the calorimeter is 17 radiation lenghts (0.6 nuclear interaction lengths).
Hamamatsu R-7600-00-M4 multianode photomultipliers are the core sensors of the ECAL, while the coupling 
to fibers is realized by means of plexiglass light guides, that maximize light collection and reduce cross-talk. 
Optical contact is enhanced by silicone joints, placed on light guides. 
Each PMT accommodates four 8.9 $\times$ 8.9 mm$^2$ anodes. Anodes define ECAL granularity: the detector allows for 18 longitudinal and 72 lateral samplings, for 
a total of 18 $\times$ 72 = 1296 readout units, usually referred to as \emph{cells}. A 3-D imaging of the shower development has been achieved by alternating 5 \emph{superlayers}
(SL), i.e. 10 layers, with fibers along X axis and 4 SL, i.e. 8 layers with fibers along Y axis.\\
In order to obtain the necessary energy resolution on Minimum Ionizing Particles (used for detector performance monitoring and equalization) as well as 
to measure energies up to 1 TeV, the digitization is performed at two different gains: High Gain for low energy measurements, and Low Gain for highest ones, with a conversion factor HG/LG of about 33.
Besides the 8 signals from anodes, each PMT last dynode signal is also readout and its information used to build up ECAL standalone trigger.\\
\subsection{Trigger}
The AMS-02 calorimeter has a key role in photon identification. 
When a high energy-deposit is recorded in the innermost layers of the calorimeter, a fast signal is generated and processed by the trigger board, which can enable the event acquisition, within 240 ns.\\
In order 
to discriminate between photons and charged particles, entering the calorimeter from outside its field of view, an angular cut is performed as an additional trigger request.
The incident particle direction is evaluated by taking, for each view, the average position of the fired PMTs and a cut at 20 degrees is applied to select particles inside
AMS-02 geometrical acceptance.\\
The efficiency for photons is about 20$\%$ at 1 GeV 
and reaches about 99$\%$ at 10 GeV, for an average polar orbit rate of 115 Hz (approximately 10$\%$ of total AMS-02 rate).
\section{Detector performance }
\begin{figure}[ht!]
 \centering
 \includegraphics[width=0.45\textwidth,clip]{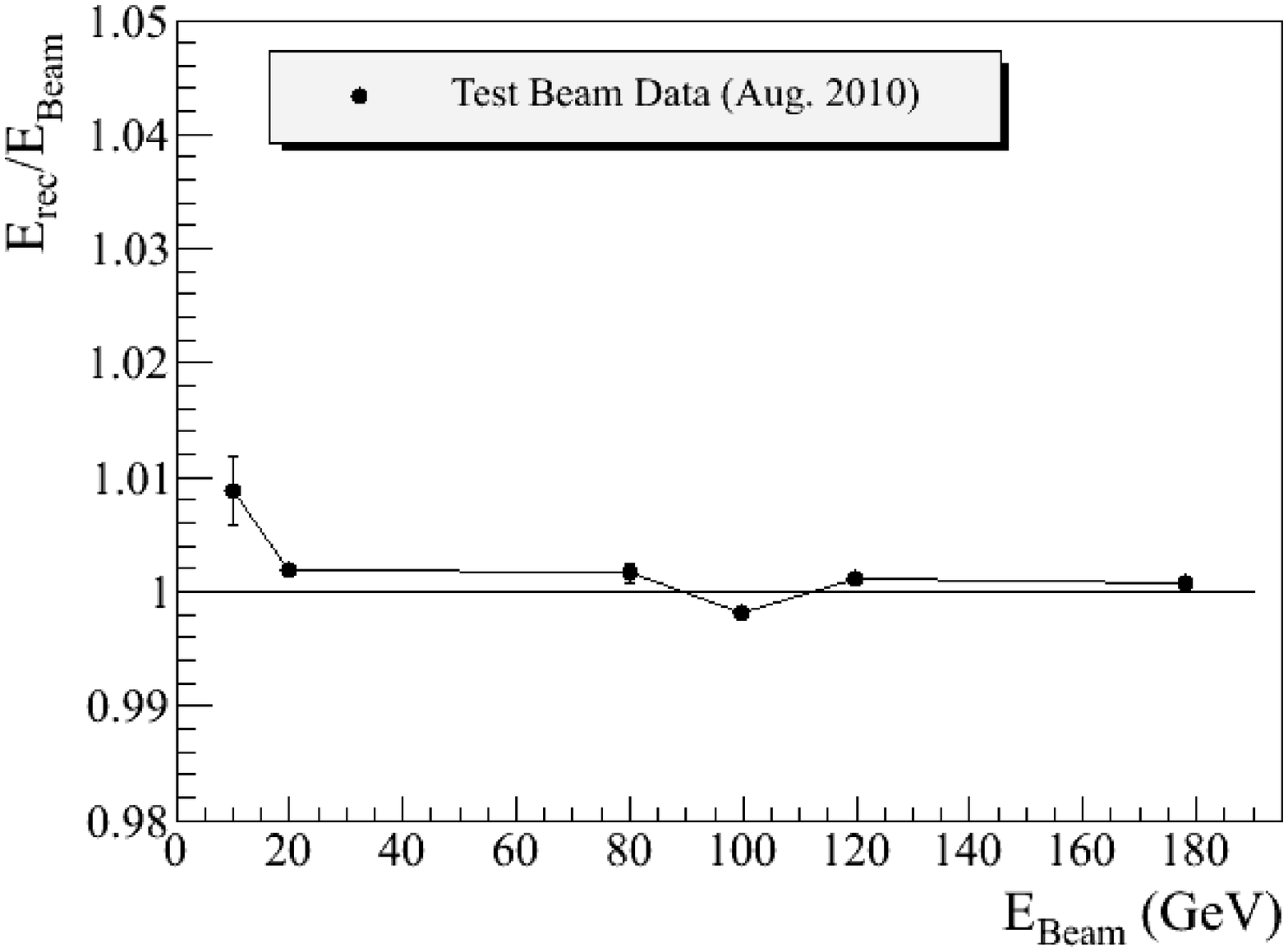}%
 \includegraphics[width=0.47\textwidth,clip]{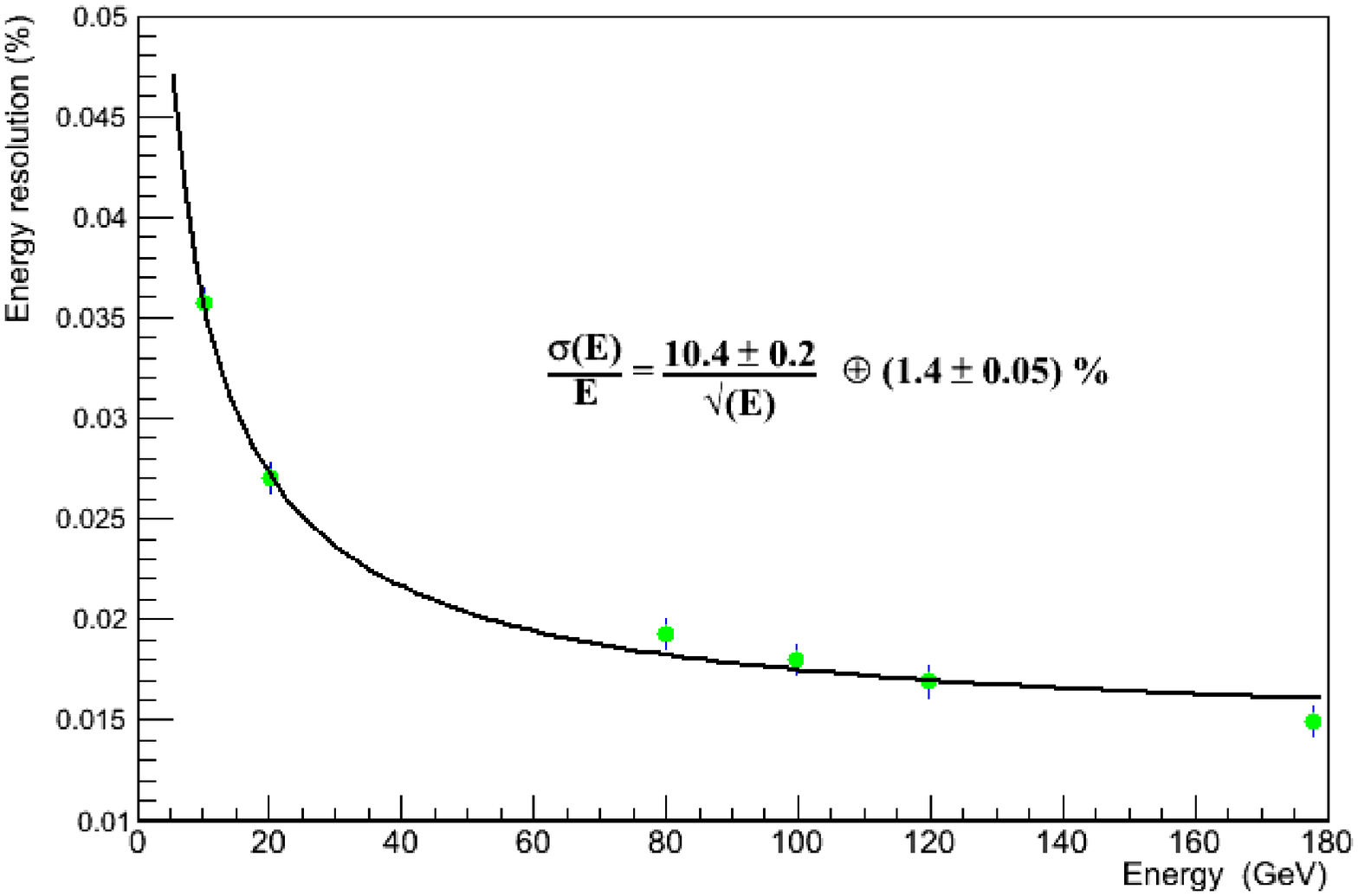}      
  \caption{{\bf Left:} Linearity as a function of the energy beam. {\bf Right:} Energy resolution as a function of the energy beam.}
  \label{vecchi:fig3}
\end{figure}

The full AMS-02 detector has been tested in 2010 using the primary 400 GeV proton beam of the CERN Super Proton Synchrotron (SPS): protons, electrons and positrons beams of energies between 8 and 400 GeV have been used to evaluate the performance of the detector.\\
Before the energy can be reconstructed, several processes have to be taken into account: the first step is to equalize all channels, i.e. to obtain 
the same response to the same energy deposit.\\ 
After this step is completed, the deposited energy must be corrected for leakage, in order to recover the fraction of the shower energy not deposited in the calorimeter.
After these corrections, deviation from linearity
is smaller than 1$\%$ in the energy range between 8 and 180 GeV, as can be seen in the left plot of figure~\ref{vecchi:fig3}, while the 
the energy resolution, is below 2 $\%$ at energies higher than 50 GeV, as can be seen in figure~\ref{vecchi:fig3}-right.\\
Angular resolution is also an important parameter, in particular for gamma ray physics.
Under the assumption that a shower 
initiated by an electron is compatible to a shower initiated by a photon in the ECAL,
the angular resolution has been computed using electrons Test Beam data.
\begin{figure}[ht!]
 \centering
 \includegraphics[width=0.5\textwidth,clip]{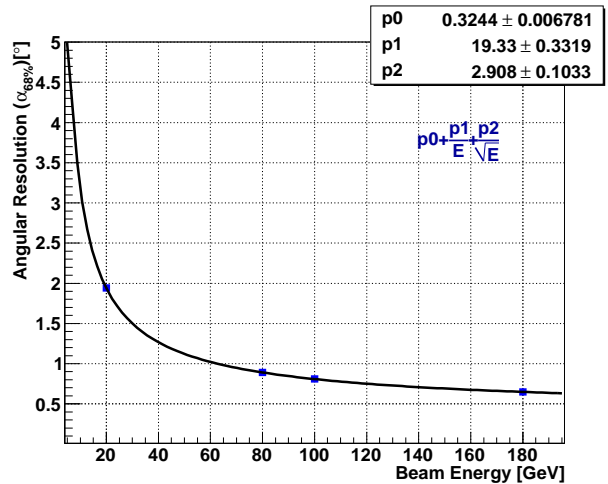}      
  \caption{ECAL angular resolution, evaluated as the angular interval
containing 68$\%$ of reconstructed angles, a function of energy, for Test Beam data. For electrons, the angle between 
the tracker track and the shower axis direction is measured.}
  \label{vecchi:fig45}
\end{figure}
Figure \ref{vecchi:fig45} shows angular resolution as a function of the beam energy: this 
result is compatible to the one obtained with flight data.
\subsection{Detector calibration in space }
The detector has been operating onboard the ISS since May 2011: a Low Earth Orbit is followed, at about 400 km 
of altitude from the Earth surface. 
The thermal environment in space is severe: Sun light as well as the position of the solar panels onboards the 
ISS can affect the thermal configuration of the instrument, causing performance variations or potential sources of permanent damage. 
Given the extreme conditions that can be experienced in space, the performance of the detector are continuously monitored. It 
has been possible to observe that the gain of the PMTs is slightly affected by temperature variations: the ratio between High and Low Gain decreases by 3$\times~10^{-5}$ per degree, while the pedestals can increase by half a percent per degree. The identification of MIPs is used to equalize the response of PMTs, the MIP value decreases by about 0.2 percent per degree. 
All those effects are accounted for in the offline analysis.
\section{Particles identification}
One of the main scientific targets of the AMS-02 detector is to identify the antimatter component out of the most abundant cosmic-ray components.
Positrons, for example, must be identified out of a background of protons, that is 10$^{3 \div 4}$ times more abundant than the positrons component. Main 
sources of background in the antimatter samples comes from charge 
confusion (i.e. electrons reconstructed as positrons) and from the wrong electron-hadron separation (i.e. protons identified as positrons).\\
In order to identify antimatter particles, all the AMS-02 subdetectors are involved: the Time-Of-Flight system is used to provide the trigger as well as to 
select down-going particles, the TRD is used to distinguish between protons and leptons, while the magnetic spectrometer is used to identify charge-one particles together with their sign and their rigidity.\\
Electromagnetic and hadronic showers differ in their spatial development and energy distribution in a way that can be distinguished by the calorimeter. The fraction of energy deposited in each ECAL layer allows for an effective separation between hadronic and 
electromagnetic showers: for hadrons 
\begin{figure}[ht!]
 \centering
 \includegraphics[width=0.4\textwidth,clip]{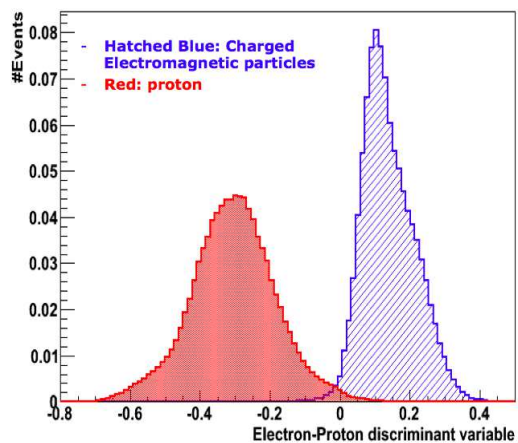}      
  \caption{Boosted Decision Tree distribution for electrons and positrons (blue distribution), as well as for protons (red distribution) for all energies.}
  \label{vecchi:fig5}
\end{figure}
of a given energy, the distribution of the fraction of energy deposited is essentially flat in each ecal layer, with 
a sharp peak at low energies for non-interacting hadrons. For electrons and positrons, the fraction of energy deposited is different from layer to layer, and 
it increases until the shower maximum is reached.\\  
Several approaches have been developed within the AMS-02 collaboration, in order to achieve the necessary separation between electrons and protons: Boosted Decision Tree, Neural Network, Likelihood methods and cut based methods. The distribution of the Boosted Decision Tree for electrons (and positrons) and hadrons is shown is figure~\ref{vecchi:fig5}.\\
Those methods, only based on ECAL-related variables, provide a rejection power on protons between 10$^2$ and 10$^3$, with a signal efficiency of 90$\%$. 
Taking into account the match between energy and momentum (as measured by the tracker), the rejection power raises up to about 10$^4$. If the TRD is also taken into account, the rejection can further improve of about one order of magnitude, while keeping the same signal efficiency. 
The rejection of those methods have been validated using Test Beam data: at energies higher than 180 GeV, where 
Test Beam data are only available for protons, Monte Carlo has been used to study their performance.
\section{Conclusions}
The AMS-02 electromagnetic calorimeter design has been described, as well as its performance. 
The main goal of the detector is to measure the energy of particles up to the TeV: the detector allows for energy linearity better than 1$\%$, and 
an energy resolution better than 2$\%$ at energies higher than 50 GeV.\\ 
Due to its granularity, ECAL allows for hadron-electron separation: a rejection power of the order of  10$^4$ is 
needed to distinguish positrons of the background of protons.
This goal is achieved using informations concerning the shower lateral and longitudinal development in the ECAL, as well 
as the match between energy and momentum (as measured by the magnetic spectrometer). The use of the TRD can further improve the rejection power, up to few hundreds GeV.


%
\end{document}